\documentclass[11pt]{article}
\newcommand{\bmu}{\mbox{\boldmath$\mu$}} 
\newcommand{\balpha}{\mbox{\boldmath$\alpha$}} 
\newcommand{\bomega}{\mbox{\boldmath$\omega$}} 
\newcommand{\bsigma}{\mbox{\boldmath$\sigma$}} 
\newcommand{\bi}[1]{\mbox{\boldmath$#1$}}
\usepackage{graphicx}

\newcommand{\cfigl}[3]{\begin{figure}[!hbtp]\centering%
 \includegraphics[width=.4\textwidth]{#2}\caption{\small{#3}}\label{#1}\end{figure}}

\begin{document} 


\title{Are the electron spin and magnetic moment\\ parallel or antiparallel vectors?}
\author{{Mart\'{\i}n Rivas}\thanks{wtpripem@lg.ehu.es}\\Dpto. de F\'{\i}sica Te\'orica,\\The University of the Basque Country,\\ 
Apdo.~644, 48080 Bilbao, Spain} 
\date{\today}

\maketitle

 \begin{abstract} 
A direct measurement of the relative orientation between the 
spin and magnetic moment of the electron seems to be never performed.
The kinematical theory of elementary particles developed by the author
and the analysis of the expectation value of Dirac's magnetic moment operator show that, 
contrary to the usual assumption, 
spin and magnetic moment of electrons and positrons might have the same relative orientation.
Two plausible experiments for their relative orientation measurement are proposed.
 \end{abstract} 
\vspace{1cm}


\vspace{2cm}
\section{Introduction}

The usual assumption concerning the magnetic dipole structure of the electron,
states that if the electron is a spinning particle of negative charge which
rotates along the spin direction then
this motion will produce a magnetic moment opposite to the spin. In the case of the positron both magnitudes,
spin and magnetic moment, will therefore have the same direction. 
But this interpretation is not supported by a classical analysis of spin, but rather by the
guess that presumably spin and angular velocity are directly related.

Dirac's analysis \cite{Dirac} of the relativistic electron shows that the spin 
and magnetic moment operators are related by
 \begin{equation}
\bmu=\frac{q\hbar}{2m}\bsigma,
 \label{eq:mus}
 \end{equation}
where $q$ is the electric charge and $\bsigma$ the spin matrix operator. For 
the electron $q=-e$, $e>0$, and therefore the spin 
and magnetic moment are antiparallel vectors while they are parallel for positrons since
$q=+e$.
However, if we make this discussion for the expectation values we obtain 
some indefiniteness in this relative orientation. Let us consider the particle analysed
in the center of mass frame. Let
\[
u_1=\pmatrix{1\cr 0\cr0\cr0\cr},\quad
u_2=\pmatrix{0\cr 1\cr0\cr0\cr},\quad
v_1=\pmatrix{0\cr 0\cr1\cr0\cr},\quad
v_2=\pmatrix{0\cr 0\cr0\cr1\cr},
\]
the usual Dirac's spinors in this frame. Spinors $u_1$ and $u_2$ are positive
energy solutions and $v_1$ and $v_2$ the negative energy ones. 
The $\sigma_z$ operator takes the form
\[
\sigma_z=\pmatrix{1&0&0&0\cr 0&-1&0&0\cr 0&0&1&0\cr 0&0&0&-1\cr},
\]
so that $u_1$ and $v_1$ are spin up states and $u_2$ and $v_2$ spin down. 
If we take the expectation value of $\mu_z$ in all these states we obtain that
for positive and negative energy states the magnetic moment has the opposite orientation
to the corresponding expectation value of the spin. 
But if the negative energy states are considered to describe the antiparticle states
then particle and antiparticle have the same relative orientation of spin 
and magnetic moment.

Now, let $C=i\gamma_2\gamma_0$ the charge conjugate operator. In 
the Pauli-Dirac representation it takes the form
\[
C=\pmatrix{0&0&0&-1\cr 0&0&1&0\cr 0&-1&0&0\cr 1&0&0&0\cr},
\]
and the charge conjugate spinors $\tilde{u}_i=Cu_i$ and $\tilde{v}_i=Cv_i$ are given by
\[
\tilde{u}_1=\pmatrix{0\cr 0\cr0\cr 1\cr},\quad
\tilde{u}_2=\pmatrix{0\cr 0\cr-1\cr0\cr},\quad
\tilde{v}_1=\pmatrix{0\cr 1\cr0\cr0\cr},\quad
\tilde{v}_2=\pmatrix{-1\cr 0\cr0\cr0\cr},
\]
so that the $\tilde{v}$ spinors represent positive energy states and $\tilde{u}$ negative
energy states of a system which satisfies Dirac's equation for a 
particle of opposite charge. Now $\tilde{u}_2$ and $\tilde{v}_2$
are spin up states and the others spin down. But for this system we have to take
for the magnetic moment operator the corresponding expression (\ref{eq:mus}) 
with $q=+e$ and therefore
the expectation values show that for positive and negative energy states of the positron
spin and magnetic moment are parallel vectors.
Then, although the expression of the magnetic moment
operator obtained in Dirac's theory is unambiguous, 
the analysis of the expectation values leads to some contradiction and 
in any case it seems that particle and antiparticle
may have the same relative orientation between both magnitudes.
We shall see in section \ref{sec:clas} that the analysis of a classical model
of electron, which satisfies when quantised Dirac's equation, leads to the same indefiniteness
in this relative orientation.

No explicit direct measurement of the relative orientation 
between spin and magnetic moment of the free electron,
known to the author, can be found in the literature 
although very high precision experiments
are performed to measure the magnitude of the magnetic moment and 
the absolute value of $g$, the gyromagnetic ratio. 
In the review article
by Rich and Wesley \cite{RichWesley} the two main methods for measuring the anomaly factor
of leptons $a=|g|/2-1$, are analysed: one kind involves precession methods which measure 
the difference between the spin precession
frequency and its cyclotron frequency in a uniform magnetic field.
The other are resonance experiments, like the ones developed by Dehmelt on a single electron
in a Penning trap \cite{Dehmelt} where the cyclotron motion, magnetron motion and axial
oscillation are monitored.
All these measurements are in fact independent of whether
spin and magnetic moment are indeed parallel or antiparallel, because they
involve measurements of the spin precession 
frequencies in external magnetic fields. 

All attempts of Stern-Gerlach type on unpolarised beams to separate 
electrons in inhomogeneous magnetic fields have failed 
and Bohr and Pauli claimed that this failure was 
a consequence of the Lorentz force on charged particles 
which blurred the splitting. Nevertheless
Seattle experiments on a single electron \cite{Dehmelt2}
show a ``continuous Stern-Gerlach type'' of interaction producing an ``axial'' oscillation
of the particle in the direction orthogonal to its cyclotron motion.
But these experiments are not able to determine the relative orientation 
between these magnitudes. Batelaan et al. \cite{Batelaan}, propose an alternative 
device which according to Dehmelt's suggestions minimize
the Lorentz force by using an external magnetic field along the electron velocity
but maximize the spin force by using large magnetic field gradients in that direction. 
They obtain numerically a polarization of the electron beam
along the direction of motion. Perhaps an experimental setup in these terms will be able to clarify
the relative orientation between these two vector magnitudes.

After giving in the next section a short review to the kinematical formalism of 
spinning particles 
developed by the author, in section {\ref{sec:directexperi}} a 
direct experiment and in \ref{sec:indirexper} an indirect 
experiment will be suggested to check the relative orientation 
of spin and magnetic moment.

\section{Clasical spinning particles}
\label{sec:clas}

The kinematical theory of elementary spinning particles \cite{Rivas1}
produces a classical description of spin and an elementary
particle in this formalism is a pointlike object. In this point we locate the charge of the particle
and its motion can be interpreted as a translational motion of its center of mass
and a harmonic motion of the center of charge around the center of mass.
Once the spin direction is fixed, the motion
of the point charge is completely determined. 
If we consider as {\em the particle} the positive energy
solution and of negative electric charge, 
then the spin and magnetic moment for 
both the electron and positron are 
described by parallel vectors. If we consider that the particle has positive
charge we get the opposite orientation. We thus obtain the same indefiniteness
as in the previous analysis of the expectation value of Dirac's magnetic moment operator.

Let us review the main highlights of the mentioned approach:
\begin{itemize}
\item The classical variables that characterise the initial and final state
of a classical elementary spinning particle in a Lagrangian approach are precisely
the variables used as parameters of the kinematical group of space-time symmetries 
or of any of its homogeneous spaces. 
Any element of the Poincar\'e group can be parametrised in terms of the time and
space translation and the relative velocity and orientation 
among inertial observers. Therefore,
a relativistic spinning particle is described by the variables
time $t$, position ${\bi r}$, velocity ${\bi v}$ and orientation $\balpha$.
We shall call to these variables the kinematical variables and the manifold they span
the kinematical space of the system.

\item A classical spinning particle is thus described as a point with orientation. 
The particle moves and rotates in space. Point ${\bi r}$ describes its position in space
while $\balpha$ describes its spatial orientation.
But what point ${\bi r}$ describes is the position of the charge, 
which is in general a different 
point than its center of mass ${\bi q}$, and in general 
${\bi r}$ describes a harmonic motion around ${\bi q}$,
usually called this motion the {\em zitterbewegung}. 

\item When expressed the Lagrangian in terms of the kinematical variables 
it becomes a homogeneous function of first degree
of the derivatives of the kinematical variables and consequently 
it also depends on $\dot{\bi v}$, the acceleration of point ${\bi r}$, and on 
$\dot{\balpha}$ or equivalently on the angular
velocity $\bomega$. It turns out that it can be written as
\begin{equation}
L=T\dot{t}+{\bi R}\cdot\dot{\bi r}+{\bi V}\cdot\dot{\bi v}+{\bi W}\cdot\bomega,
\label{eq:L}
\end{equation}
where $T=\partial L/\partial\dot{t}$, ${\bi R}=\partial L/\partial\dot{\bi r}$,
 ${\bi V}=\partial L/\partial\dot{\bi v}$ and  ${\bi W}=\partial L/\partial\bomega$.

\item For a free relativistic particle, when analyzing the
invariance under the different one-parameter subgroups of the Poincar\'e group, 
Noether's theorem 
determines the usual constants of the motion which take the following form 
in terms of the above magnitudes: Energy,
\[
H=-T-{\bi v}\cdot\frac{d{\bi V}}{dt},
\]
linear momentum, 
\begin{equation}
{\bi P}={\bi R}-\frac{d{\bi V}}{dt},
\label{eq:p}
\end{equation}
kinematical momentum
\begin{equation}
{\bi K}=\frac{H}{c^2}{\bi r}-{\bi P}t-\frac{1}{c^2}{\bi S}\times{\bi v},
\label{eq:k}
\end{equation}
and angular momentum
\begin{equation}
{\bi J}={\bi r}\times{\bi P}+{\bi S},
\label{eq:j}
\end{equation}
where the observable ${\bi S}$, takes the form
\begin{equation}
{\bi S}={\bi v}\times{\bi V}+{\bi W}.
\label{eq:s}
\end{equation}

\item The linear momentum (\ref{eq:p}) is not lying along the velocity 
${\bi v}$ of point ${\bi r}$. Point ${\bi r}$
does not represent the center of mass position. If in terms of the last term in (\ref{eq:k}) 
we define the position vector
\[
{\bi k}=\frac{1}{H}\,{\bi S}\times{\bi v},
\]
then the center of mass position can be defined as ${\bi q}={\bi r}-{\bi k}$, 
such that the kinematical momentum can be written as
\[
{\bi K}=\frac{H}{c^2}{\bi q}-{\bi P}t.
\]
Taking the time derivative of this expression leads for the linear momentum to
\[
{\bi P}=\frac{H}{c^2}\,\frac{d{\bi q}}{dt},
\]
which is the usual relativistic expression of the linear 
momentum in terms of the center of mass velocity.
Observable ${\bi k}$ is the relative position of 
point ${\bi r}$ with respect to the center of mass  ${\bi q}$. 

\item The observable ${\bi S}$ is the classical equivalent of Dirac's spin operator, because
it satisfies the free dynamical equation
\[
\frac{d{\bi S}}{dt}={\bi P}\times{\bi v},
\]
which does not vanish because ${\bi P}$ and ${\bi v}$ are not parallel vectors
and where the velocity operator in Dirac's theory becomes ${\bi v}=c\balpha$ in 
terms of Dirac's $\balpha$ matrices.

\item The structure of the spin observable is twofold. One, ${\bi v}\times{\bi V}$, 
is related to the zitterbewegung
or charge motion around its center of mass and another ${\bi W}$ to the rotation of the particle.

\item The magnetic moment is produced by the charge motion and is thus related 
only to the zitterbewegung part of the spin. It is because the spin has 
another contribution coming from the rotation
that a pure kinematical interpretation of the gyromagnetic ratio 
\cite{Rivas3} has been given.

\item The classical system that when quantised satisfies Dirac's equation \cite{Rivas2} corresponds
to a particle whose charge is moving at the speed of light, and therefore if $v=c$ is constant,
the acceleration is always orthogonal to the velocity.

\item If we take in (\ref{eq:k}) the time derivative of this expression 
and afterwards its scalar product with vector ${\bi v}$, since $v=c$ we get the relationship
\[
H-{\bi P}\cdot{\bi v}-\frac{1}{c^2}\left({\bi S}\times\frac{d{\bi v}}{dt}\right)\cdot{\bi v}=0
\]
which is the classical equivalent of Dirac's equation.

\item The center of mass observer is defined by the conditions ${\bi P}=0$ and ${\bi K}=0$.
For this observer we see from (\ref{eq:j}) 
that the spin ${\bi S}$ is a constant of the motion. If the system has positive energy $H=+mc^2$,
from (\ref{eq:k}) we get
\[
m{\bi r}=\frac{1}{c^2}{\bi S}\times{\bi v},
\]
so that the charge of the particle is describing circles in a plane orthogonal to ${\bi S}$
as depicted in part a) of figure \ref{fig:1}. 
Part b) is the time reversed motion
of this particle which corresponds to its antiparticle or to 
a particle that in the center of mass frame has energy $H=-mc^2$.
If the particle is negatively charged then particle and antiparticle have their magnetic moment 
along the spin direction. If we consider as the particle the positively charged one
we obtain the opposite orientation for the magnetic moment. In any case the magnetic moment of the particle and antiparticle
have the same relative orientation with respect to the spin.

\cfigl{fig:1}{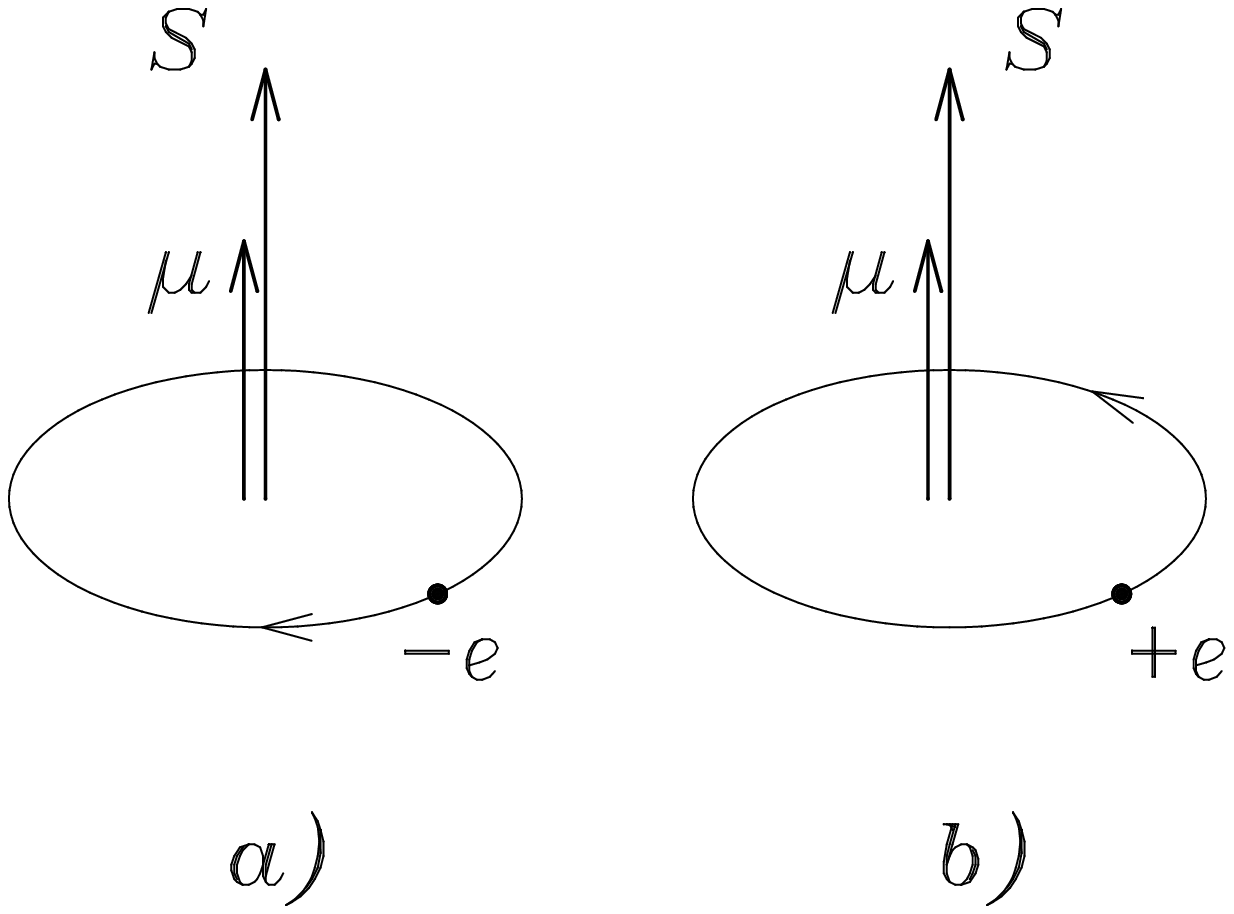}{Charge motion of the electron a) and positron b) in the center of mass frame.}

The radius 
of this motion is
\[
R_o=\frac{S}{mc}=\frac{\hbar}{2mc}= \frac{1}{2}\lambda_C=3.8\times10^{-11}\;\;{\rm cm}.
\]
The frequency of this motion is
\[
\omega_o=\frac{mc^2}{S}=\frac{2mc^2}{\hbar}=1.55\times 10^{21}\, {\rm s}^{-1}.
\]

\item It turns out that although the particle is pointlike, because of the zitterbewegung 
the charge has a localized region of influence of size $2R_o$, which is Compton's wave length.
The latest LEP experiments at CERN establish an upper bound of $10^{-17}$cm for the radius 
of the charge of the electron,
which is consistent with this pointlike interpretation, 
while its quantum mechanical behaviour is produced 
for distances of its Compton wave length, six orders of magnitude larger.

\item Properly speaking what this formalism shows is that 
the magnetic moment $\bmu$ of the electron is not an 
intrinsic property like the charge. 
It is produced by the motion of the charge and therefore is orthogonal
to the zitterbewegung plane. But at the same time, in the center of mass frame, 
the electron has an oscillating electric dipole
of magnitude $eR_o$ lying on the zitterbewegung plane. 
Its time average value vanishes and in low energy interactions the effect 
of this electric dipole is negligible but in high energy physics we have to take into account
the detailed position of the charge and thus the electric dipole contribution
is not negligible. This electric dipole is not related
to a loss of spherical symmetry of some charge distribution. The charge distribution
is spherically symmetric because it is just a point. This dipole 
is just the instantaneous electric dipole moment
of the charge with respect to the center of mass.

\end{itemize}

As an approximation we can consider the classical electron 
as a point, its center of mass, where we also locate the charge. But at the same time
we have to assign to this point two electromagnetic properties,
a magnetic moment lying along or opposite to the spin
direction and an oscillating electric dipole, of frequency $\omega_o$, 
on a plane orthogonal to the spin.

A more detailed analysis of the dynamics gives rise to the dynamical equations for the center of
mass ${\bi q}$ and center of charge ${\bi r}$ in an external electromagnetic field as given by \cite{Rivas4}: 
 \begin{eqnarray}
m\ddot{\bi q}&=&\frac{e}{\gamma(\dot{q})}\left[{\bi E}+\dot{\bi r}\times{\bi B}-\dot{\bi q}\left(\left[{\bi E}
+\dot{\bi r}\times{\bi B}\right]\cdot\dot{\bi q}\right)\right],\label{eq:mq2}\\
\ddot{\bi r}&=&\frac{1-\dot{\bi q}\cdot\dot{\bi r}}{({\bi q}-{\bi r})^2}\left({\bi q}-{\bi r}\right).\label{eq:r2enq}
 \end{eqnarray}
Here, an over dot means a time derivative and the external fields are defined at the charge position ${\bi r}$,
and it is the velocity of the charge that produces the magnetic force term.

\section{A direct measurement}
\label{sec:directexperi}

Since the classical model depicted in figure {\ref{fig:1}} satisfies 
when quantised Dirac's equation \cite{Rivas2}
it is legitimate to use it for analysing the interaction with an external electromagnetic field
from a classical viewpoint. We can alternatively use the Bargmann-Michel-Telegdi equation
for the spin evolution\cite{BMT}, but this approach assumes a minimal 
coupling for the charge and also an anomalous 
magnetic moment coupling. However in our approach since the spinning electron is a point charge
such that the magnetic moment is a consequence of the zitterbewegung, we only have to consider
a minimal coupling prescription which is closer to quantum electrodynamics
in which no anomalous magnetic moment coupling for the electron is present.

The proposed experiment is to send a beam of transversally polarised electrons
or positrons and check the interaction of their magnetic moment with an external 
magnetic field. 
As suggested by Batelaan et al. \cite{Batelaan}
we shall consider a region of low or negligible magnetic field but with a 
non-negligible field gradient such that the deflection of the beam 
is mainly due to the magnetic dipole structure. 

We consider a beam moving along the positive direction of 
the $OY$ axis and with the transversal spin pointing along the positive $OZ$ axis.
The external magnetic field will be produced by two conducting wires parallel to the $OY$ axis, 
separated by a distance $2b$ and contained
in the $YOZ$ plane of a cartesian frame (see figure \ref{fig:2}). 
If they carry a current in the same direction, 
then the magnetic field vanishes along the $OY$ axis and is very low in its neighborhood. 
The square depicted in the figure represents the region, using for computation,
where the initial position of the center of mass of the electrons in the beam is contained. 
We have found no experimental
evidence of one such a device which could be useful to analyse the magnetic moment
of free charged particles as an alternative to the Stern-Gerlach magnets 
which do not work properly with charged particles. 

\cfigl{fig:2}{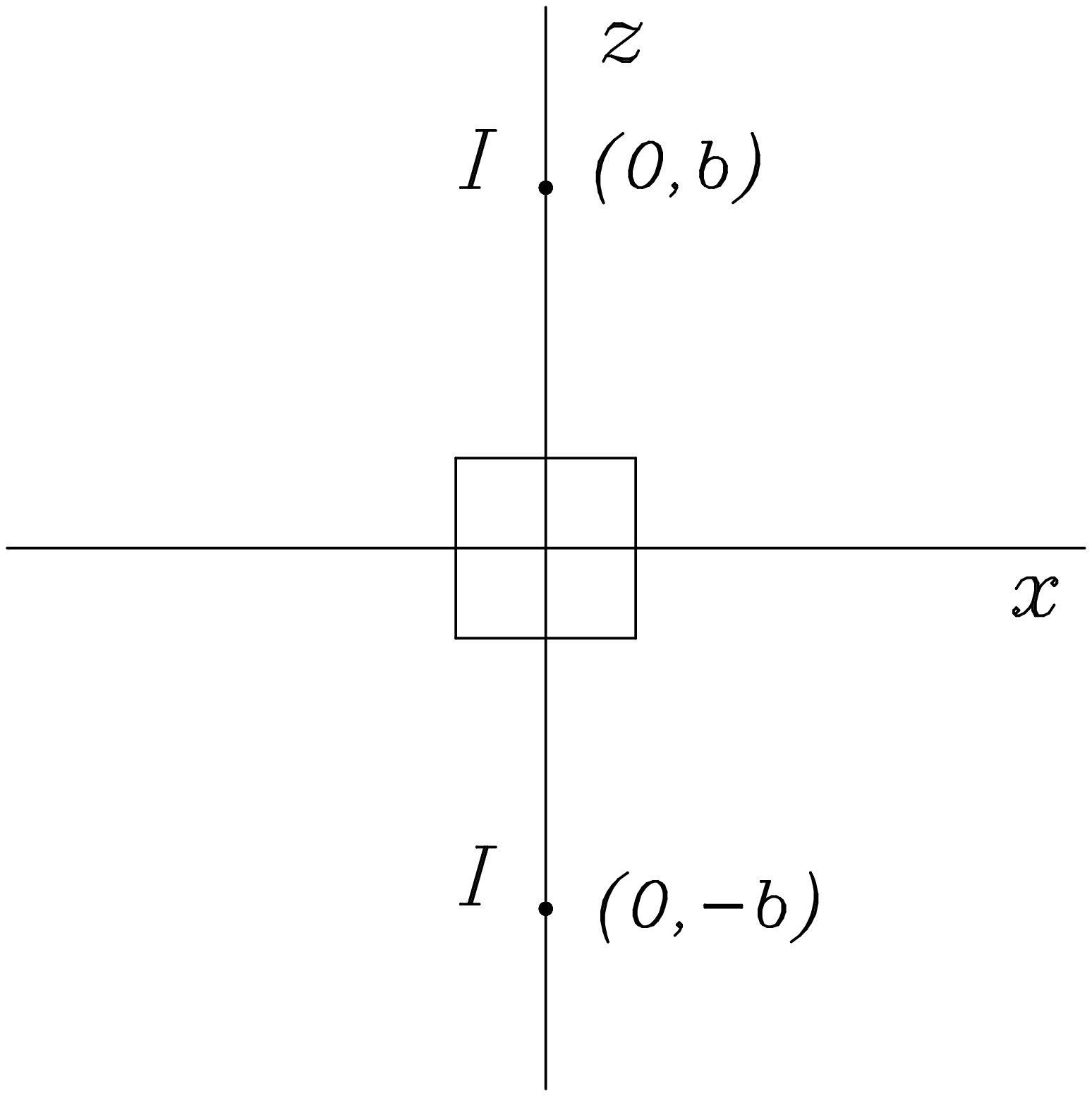}{A transversally polarised electron beam of 
square cross section is sent along the $OY$ axis 
into the magnetic field created by two conducting wires, separated 
a distance $2b$, perpendicular to 
the figure and carrying a current $I$ in the same direction.}

The magnetic field produced by the current takes the form:
 \begin{eqnarray*}
B_x&=&\frac{k(z+b)}{x^2+(z+b)^2}+\frac{k(z-b)}{x^2+(z-b)^2}, \quad 
B_y=0,\\
B_z&=&-\frac{kx}{x^2+(z+b)^2}-\frac{kx}{x^2+(z-b)^2},
 \end{eqnarray*}
where $k=I/2\pi\epsilon_0c^2$, $I$ is the intensity of the current 
and $\epsilon_0$ the permittivity of the vacuum. 

To compute numerically the motion of a polarised electron beam in an external 
magnetic field, we shall use the above dynamical equations (\ref{eq:mq2}) and (\ref{eq:r2enq})
with initial conditions such that the zitterbewegung plane of each electron 
is the $XOY$ plane and the charge motion
produces a magnetic moment pointing along the positive $OZ$ axis. 
For the center of mass position
we shall consider the electrons 
uniformly distributed in the mentioned square region.

\cfigl{fig:3}{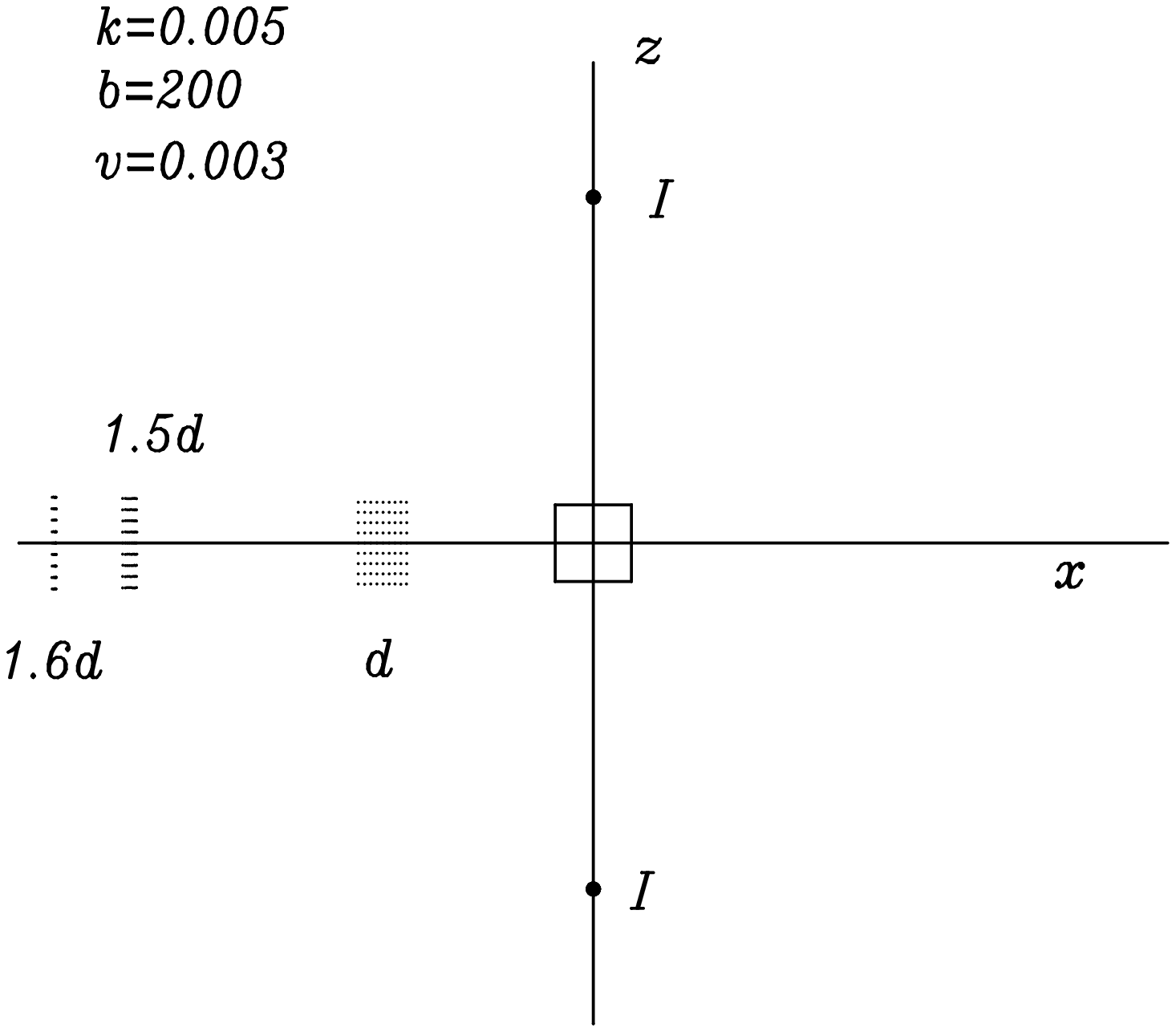}{Position of the beam after travelling a distance $d$, $1.5d$ and $1.6d$
along the $OY$ direction in the magnetic field created by the two parallel currents. The separation
between the wires is $200\lambda_C$, the velocity of the center of mass of electrons is $0.003c$ 
and the current $I=3$A.}

In figure \ref{fig:3} we depict the situation of the polarised electron beam after travelling some distance
inside the magnetic field region. The dots represent the center of mass position of a sample of particles 
which in the incoming beam are distributed uniformly in the shown square region. 
With the magnetic moment of the
electrons pointing upwards we see a deflection (and also a focusing effect) 
to the left. The deflection is of the same amount to the right
for electrons with the magnetic moment pointing down. It is checked that the deflection is independent
of the initial position of the electron charge compatible with the initial center of mass position.
We also obtain an equivalent deflection for many other values of the separation between wires, 
current and initial beam velocity.

We thus expect from this experiment that if the beam is deflected to the left
then spin and magnetic moment are parallel vectors, while they are antiparallel for right deflection. 
The interaction does not modify the spin orientation so that we can check the polarization
of the beam at the exit by some direct method of electron absorption like the one devised 
for measuring the spin of the photon in circularly polarised light beams \cite{Beth}.

Although this device is considered for analysing charged particles the deflection
is produced in the low magnetic field region so that it
is mainly due to the interaction with the magnetic moment. This is one of the reasons
to consider this device as an alternative to the Stern-Gerlach magnets
to separate unpolarised beams.

\section{An indirect measurement}
\label{sec:indirexper}

As an indirect experiment we shall measure the relative
orientation of spin and magnetic moment of electrons bounded in atoms. 

Let us consider a material system formed by atoms of some specific
substance. Let us send a beam of circularly polarised light of such an energy 
to produce electron transitions on these atoms from an S-state ($l=0$ orbital angular momentum) 
into another S-state. 
Let us assume that the photons of the circularly polarised beam have their spins pointing forward.
In this case the transition only affects to the electrons with the spin pointing backwards
such that after the transition the excited electrons have their spins in the forward direction.
If we now introduce a magnetic field in the forward direction to observe the Zeeman splitting then
the measurement of the additional interaction energy $-\bmu\cdot{\bi B}$ will give us only one of the 
two expected transition lines of the emission spectrum from which we determine the relative orientation
between $\bmu$ and ${\bi B}$ and therefore between $\bmu$ and ${\bi S}$.

\section*{Acknowledgments}

I like to acknowledge the use of the computer program {\em Dynamics Solver} developed
by Aguirregabiria \cite{Aguirre} with which all the numerical computations contained in this 
paper have been done. I am very much indebted to Aguirregabiria's help.

I also acknowledge the suggestion by Prof. Dieter Trines of DESY, Hamburg, during
the Prague SPIN-2001 congress, that leads to the proposed indirect measurement experiment.

This work has been partially supported by the Universidad del Pa\'{\i}s Vasco/Eus\-kal Herriko Unibertsitatea 
under contract UPV/EHU 172.310 EB150/98 and by General Research Grant UPV 172.310-G02/99.

 \end{document}